\documentclass[twocolumn,showpacs,preprintnumbers,amsmath,amssymb,prb]{revtex4}

\usepackage{color}
\usepackage{graphicx}
\usepackage{bm}
\usepackage{natbib}

\newcommand{\eg}{{\it e.\,g.}}
\newcommand{\ie}{{\it i.\,e.}}

\begin{document}

\title{Long-term Dynamics of 
the Electron-nuclear Spin System of a Semiconductor
Quantum Dot}

\author{I. A. Merkulov}
\affiliation{Material Science
and Technology Division, Oak Ridge National Laboratory, Tennessee 37831 USA}
\affiliation{A. F. Ioffe Physico-Technical Institute, Russian Academy of Sciences, 194021 St. Petersburg, Russia}
\author{G. Alvarez}
\affiliation{Computer Science \& Mathematics 
Division and Center for Nanophase Materials Sciences, Oak Ridge National Laboratory, \mbox{Oak Ridge, TN 37831} USA}
\author{D. R. Yakovlev}
\affiliation{Experimental
Physics 2, TU Dortmund University, 44221 Dortmund, Germany}
\affiliation{A. F. Ioffe Physico-Technical Institute, Russian Academy of Sciences, 194021 St. Petersburg, Russia}
\author{T. C. Schulthess}
\affiliation{Institute for Theoretical Physics and Swiss National Supercomputer Center, ETH
Zurich, Wolfgang Pauli Strasse 27, 8093 Zurich, Switzerland
}


\begin{abstract}
A quasi-classical theoretical description of polarization and relaxation
of nuclear spins in a quantum dot with one resident electron
is developed for arbitrary mechanisms of electron spin polarization.
The dependence of the electron-nuclear spin dynamics on the correlation time $\tau_c$ of
 electron spin precession, with frequency $\Omega$, in the nuclear hyperfine field is analyzed.
It is demonstrated that the highest nuclear polarization is achieved for a correlation
time close to the period of electron spin precession in the nuclear field.
For these and larger correlation times, the indirect
hyperfine field, which acts on nuclear spins, also reaches a maximum. This maximum
is of the order of the dipole-dipole magnetic field that nuclei create on each other. 
This value is non-zero even if the average electron polarization vanishes.
It is shown that the transition from short correlation time to $\Omega\tau_c\gtrsim1$
does not affect the general structure of the equation for nuclear spin temperature 
and nuclear polarization in the Knight field, but changes the values of parameters, which now 
become functions of $\Omega\tau_c$. For correlation times larger than
the precession time of nuclei in the electron hyperfine field, it is found that three thermodynamic 
potentials ($\chi$, $\bm{\xi}$, $\varsigma$) characterize 
the polarized electron-nuclear spin system.  The values of these potentials are
calculated assuming a sharp transition from short to long correlation times, and the relaxation mechanisms of these potentials
are discussed. The relaxation of the nuclear spin potential is simulated numerically showing that
high nuclear polarization decreases relaxation rate.

\end{abstract}
\maketitle
\section{Introduction}

The electron-nuclear spin system (ENSS) of a semiconductor
quantum dot (QD) has been under intensive investigation in recent
years\cite{re:dyakonov08,re:henneberger09,re:awschalom02}.
This strong interest has been motivated by potential spintronics
and quantum information applications, for which semiconductor quantum
dots are promising\cite{re:henneberger09,re:awschalom02}. 
The spin dynamics of this
system is described by a variety of relaxation times which range from
nanoseconds to seconds. 

Optical orientation is a commonly used method to create and control\cite{re:meyer84} the ENSS
with a high degree of polarization. Nuclear polarization is caused by the Fermi hyperfine
interaction\cite{re:fermi30} between nuclear spins and photo oriented electrons. A simple
theoretical description of the ENSS behavior is
the short correlation time approximation (SCTA). (The electron correlation time, $\tau_c$, 
is the
characteristic time of the free coherent undisturbed electron spin
precession in the hyperfine field of the nuclei.) The SCTA is valid
if the frequencies of electron spin precession (${\Omega }$) in the local nuclear
hyperfine field,  and nuclear spin
precession ($\omega$) in the electron hyperfine field are small enough: $\Omega\tau_c\ll1$,
and $\omega\tau_c\ll1$. In a quantum dot, the electron interacts with a macroscopic
number, $N$, of nuclei, \ie, $N\propto 10^5$ 
and $\Omega\gg \omega$.
It follows that 
the SCTA can be used if $\Omega\tau_c\ll1$. 

When $\Omega\tau_c\ll1$ holds,
frequencies of electron
and nuclear spin precession are practically constant during the time $\tau_c$.
Small deviations of
these frequencies during $\tau_c$ 
is a small perturbations to the
spin motion. This deviation is the mechanism for the slow transfer of spin
polarization between electron and nuclei. 

The SCTA is
valid in many experimental scenarios. 
However, studying the problem  beyond the SCTA stimulates  experimental and theoretical
investigations in the regime of intermediate correlation time,
where $\Omega\tau_c\gtrsim 1$, and 
$\omega\tau_c\ll 1$, and in the regime of long correlation time, where $\omega\tau_c\gg1$.
 These regimes occur at low temperature, and under constant wave (CW) light of low intensity or in the
 darkness, respectively. The intermediate
regime may be realized also by using circularly polarized light pulses.
In this case, short periods of light illumination and high
photoelectron concentration are alternated with long periods of
darkness, when the ENSS motion is undisturbed.

In this paper we discuss the behavior of an ensemble of quantum
dots, each containing a single resident electron. We consider
the simplest experimental
scenario, where in the first step some external action (\eg, circularly polarized photons) orients the
resident electron which polarizes the QD nuclei, and in the second step the
ENSS dynamics is determined only by interactions between quantum dot
spins. The electron becomes polarized due to the
spin exchange with optically oriented photo
carriers. After switching off the light, a relaxation takes place.
It is characterized by a long relaxation time $T\gg\omega^{-1}$.
The relaxation  is a result of 
the dipole-dipole
interaction between neighboring nuclei, and of the electron-phonon
interaction\cite{re:abragam96,re:erlingsson01,re:khaetskii00}.

In Section~\ref{sec:light} we demonstrate 
(i) that the maximum rate of
nuclear polarization by optically oriented electrons is reached
for  $\Omega \tau_c\approx 1$, (ii) that
the  nuclear polarization is a result of the nuclei cooling by
spin oriented electrons in the Knight field (connected with the time averaged
electron polarization), and (iii) that the spatial dependence of the hyperfine interaction
decreases the photoinduced nuclear
polarization. In the intermediate regime  
 $\Omega \tau_c$ mostly influences  the nuclear spins relaxation times. In
Section~\ref{sec:darkness}, we discuss the difference in the ENSS description between the intermediate
and long correlation time approximations.  We also present
numerical results  for the dipole-dipole
relaxation. These calculations show an increasing
relaxation time with increasing nuclear polarization. All cases
are calculated for a Spherical Quantum Dot with Infinitely high Barrier
(SQDIB), which allows us to
ignore the exponentially small escape of the electron wave
function out of the quantum dot. We limit the description of the ENSS
interaction with the QD's environment by introducing the leakage factor
approximation, and we do not discuss the specifics
of spin diffusion of the nuclear polarization outside the 
QD\cite{re:zhang06,re:erlingsson04,re:yubashyan05}.

To model the spin system we use the
quasi-classical approximation, which is valid for quantum dots with
large numbers of nuclear spins\cite{re:merkulov02,re:hen07}. Finally, in section~\ref{sec:summary} we summarize our main results.

\section{Polarization of the electron-nuclear spin system in a quantum dot}\label{sec:light}

In a semiconductor quantum dot with one resident electron
the hyperfine interaction creates a localized electron-nuclear spin
polaron\cite{re:degennes60,re:furdyna88b,re:furdyna88}. 
In this section we discuss this system's behavior in the limit
of short and intermediate correlation time, $\omega\tau_c\ll1$. In these regimes the
frequency of the nuclear spin precession in the electron hyperfine field
is lower than the frequency of external perturbations of electron field orientation. We
do not specify the character of the external interaction, but will simply 
suppose that this interaction can partially orient the electron spin.
The exchange scattering 
of an optically polarized
carrier is an example of such an external interaction.
 
\subsection{Electron spin precession in the nuclear hyperfine field}

In GaAs-like semiconductors the electron and nuclear spins 
are coupled by the Fermi hyperfine interaction,
\begin{equation}
\hat{H}=\frac{\pi}{3}{\mu}_B\sum_n
\frac{\mu_n}{I_n}(\bm{s}\cdot
\bm{I}_n)\delta(\bm{r}-\bm{R}_n),
\label{eq:hyperfine}
\end{equation}
where $\mu_B$ is the Bohr
magneton, $\bm{s}$  and $\bm{r}$  are
the spin and position of the electron, 
$\mu_n$, $\bm{I}_n$  and
$\bm{R}_n$ are the magnetic moment, spin and
position of the n-th nucleus, respectively. The sum in Eq.~(\ref{eq:hyperfine}) runs 
over all the nuclei inside the QD. The hyperfine energy has a maximum 
when $\bm{s}$ and $\bm{I}_n$ are parallel, and a minimum when they
are anti-parallel. In the following, and for simplicity, we will suppose
that all nuclei have the same spins and magnetic moments
$I_n\equiv I$ and $\mu_n\equiv \mu_I$.

A QD contains a macroscopic number, $N$, of nuclear spins, $N\propto10^5\gg1$. 
Therefore, the frequency of 
electron spin precession in the nuclear hyperfine field,
\begin{equation}
\bm{\Omega} =\sum_n\omega_n\bm{I}_n,
\label{eq:precess-hyperfine}
\end{equation} 
is distributed in a wide region from zero to
$\Omega_{\rm max}$ given by 
\begin{equation}
\Omega_{\rm max}=I\sum_n\omega_n.
\end{equation} 

It is usual to separate $\bm{\Omega}$ in two parts,
average and fluctuation: $\bm{\Omega} =\langle \bm{\Omega} \rangle
+\Delta\bm{\Omega}$. In the
following, average and fluctuation refer to the time evolution
and time fluctuation of the
nuclear spins. Because $\langle \bm{\Omega} \rangle$ and $\Delta \bm{\Omega}$ are not correlated,
 $\langle \Omega ^2\rangle =\langle
\Omega \rangle ^2+\langle(\Delta\Omega) ^2\rangle$.

$\bm{\Omega}$ is many orders of magnitude larger than
the frequency of nuclear spin precession in the electron hyperfine field,
 $\omega_n\bm{s}$, \ie, $\Omega\ge\Omega_{\rm fluc}$, where
\begin{eqnarray}
\Omega_{\rm fluc} & = &\sqrt{\|I\|^2
\sum_n\omega_n^2}\\\nonumber
& = &\|I\|\sqrt{\langle
\omega^2\rangle}\sqrt{N}\gg\|s\|
\sqrt{\langle\omega^2\rangle}.
\end{eqnarray}

Here,  $\Omega_{\rm fluc}=\sqrt{\langle \Omega ^2\rangle}|_{\Delta\Omega=0}$  is the characteristic value of the fluctuations of the electron
spin precession frequency, 
$\|I\|=\sqrt{I(I+1)}$,  and
 $\|s\|=\sqrt{s(s+1)}$, 
are the modulus of nuclear and electron spin, respectively,
\begin{equation}
\omega_n=\frac{16\pi\mu_B\mu_n}{3I_n\hbar}\|\psi(R_n)\|^2,
\end{equation}       
and  $\psi(R_n)$ 
is the electron wave-function on the n-th
nucleus\footnote{$\|\psi(R_n)\|^2$ 
and $\omega_n$  depend on the type of nucleus and on its position in the crystal cells  and  QD.}.

\begin{figure}
\centering{
\includegraphics[width=8cm]{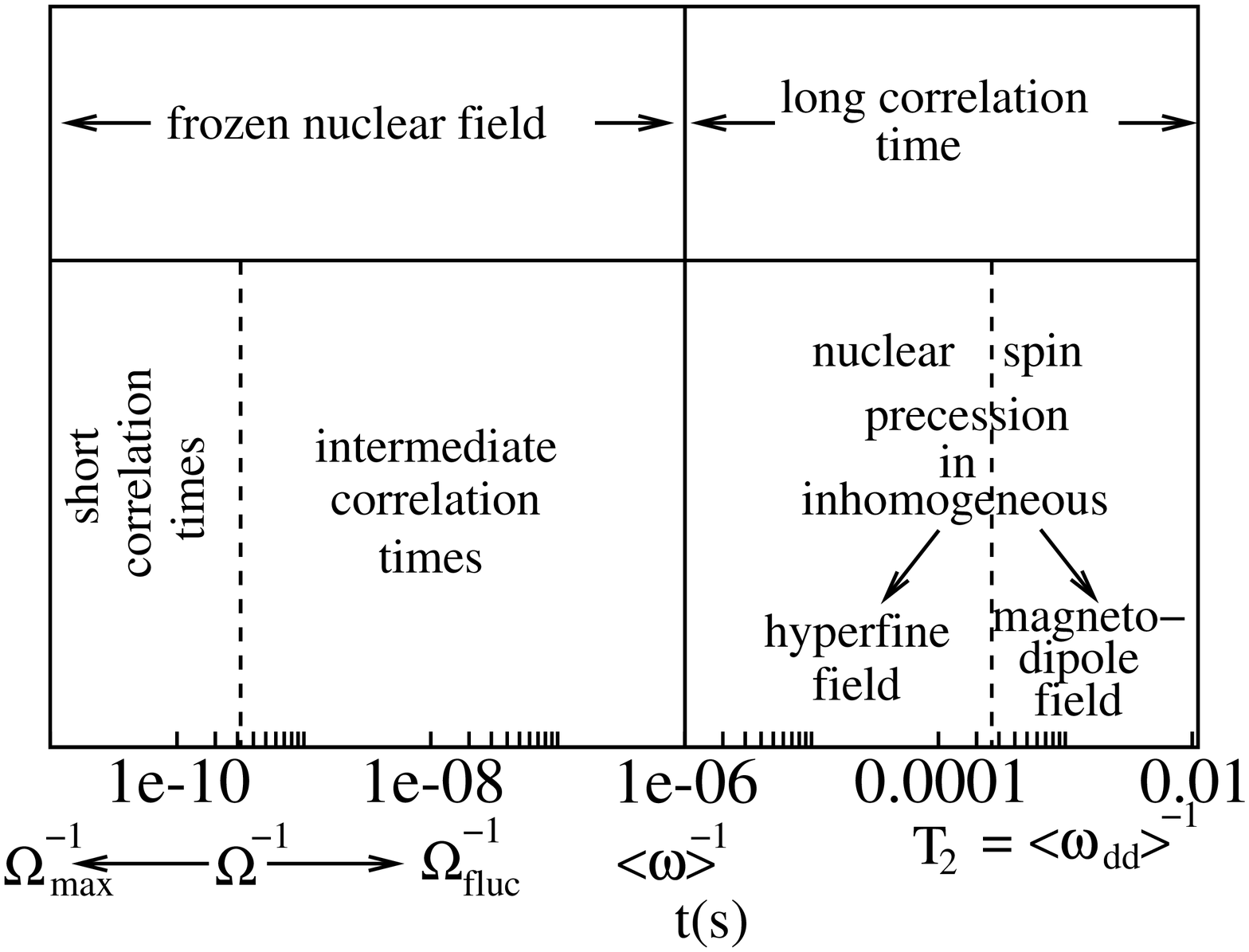}
}
\caption{\label{fig:1} Time scale for different regimes of hyperfine interaction in
quantum dot. (1) Short correlation time regime ($\tau_c<\Omega^{-1}$). During $\tau_c$ the electron
and nuclear spins rotate around a small angle. 
(2) Intermediate correlation time regime ($\Omega^{-1}<\tau_c<\langle\omega\rangle^{-1}$).
During $\tau_c$ the electron spin rotates around a large angle with constant angular velocity.
(3) Long correlation time regime ($\tau_c>\langle\omega\rangle^{-1}$). 
During $\tau_c$ the electron and nuclear spins change their direction.
For $\langle\omega_{dd}\rangle^{-1}>\tau_c>\langle\omega\rangle^{-1}$
the total nuclear spin, $I_\Sigma$ is conserved, but $\Omega$ changes 
direction as a result of the nuclear spin precession in the non-uniform
electron hyperfine field. $\langle\omega_{dd}\rangle$ is
the frequency of nuclear spin precession in the local magnetic field
of the neighboring nuclei. For $\langle\omega_{dd}\rangle^{-1}<\tau_c$,
the dipole-dipole interaction between spins of neighboring nuclei changes $I_\Sigma$.}
\end{figure}

$\Omega_{\rm max}$ does not depend on the
quantum dot's volume because the electron wave function is
normalized in this volume; it is determined solely by the chemical composition of the QD. For
example, for a GaAs QD
one can estimate\cite{re:dyakonov08} $\Omega _{\rm max}\approx  10^{-11}s^{-1}$.
Values of $\Omega_{\rm fluc}\propto\Omega_{\rm max}/\sqrt{N}$ and $\langle\omega\rangle\propto\Omega_{\rm max}/N$
depend on the QD volume. For a quantum dot with $N\propto10^5$ nuclei $\Omega_{\rm fluc}\propto 3.10^8s^{-1}$ 
and $\langle\omega\rangle\propto10^6s^{-1}$. Characteristic frequencies and times are collected
in Fig.~\ref{fig:1}.

Under sample illumination, the photocarriers and photons scattering on a quantum dot, carrier capture and photon
absorption in QD are the main mechanisms of the ENSS interaction with the environment.
As we will show in  section~\ref{subsec:precession-hyperfine}, the nuclear
spin polarization and relaxation are a consequence of this perturbation of
the electron and nuclear spin precession in the hyperfine field.
To achieve the most effective nuclear polarization,
the frequency of these collisions, 
$\omega_{\rm coll}\equiv\tau_c^{-1}$,  should be about 
$\Omega \gg\mathit\omega_n$.
Between collisions nuclear spins change their directions by a very small angle of about
$\tau_c\langle \omega \rangle /2$, which is much less than 1. In the zero-th order approximation,
the spin, $s(t)$, of the resident electron precesses in the constant (frozen) nuclear field:
\begin{eqnarray}
\bm{s}(t)=\bm{s}_{\bm{\Omega}}
(t_0) &+ & (\bm{s}(t_0)-\bm{s}_{\Omega}(t_0))\cos(\Omega\cdot \delta t)\nonumber\\
&+&\left[\bm{e}_{\bm{\Omega}}\times
\bm{s}(t_0)\right]\sin(\Omega \cdot\delta t),
\label{eq:soft}
\end{eqnarray}
where 
\begin{equation}
\bm{s}_{\bm{\Omega}}(t_0)=(\bm{s}(t_0)\cdot \bm{e}_{\bm{\Omega}})
\cdot \bm{e}_{\bm{\Omega}},
\end{equation}
$\bm{e}_{\bm{\Omega}}=\bm{\Omega} /\Omega$,
$\bm{s}(t_0)$ is the initial spin, which is determined by
collisions,
$\delta t=t-t_0$, and $t_0$ is the time when the external action polarized the resident electron.

For the continuous-wave (CW) photoexcitation, the probability $W(\delta t)$  of
electron free (undisturbed) precession during time $\delta t$ decreases exponentially with free precession time:
 $W(\delta t)=\exp(-\delta t)/\tau_c$. The spin of localized electrons (averaged over initial
polarization $\bm{s}(t_0)$  and  $\delta t$)  can be
written as:
\begin{equation}
\overline{\bm{s}}=\frac{\bm{s}_0+\left[\bm{\Omega} \times
\bm{s}_0\right] \tau_c+(\bm{\Omega} \cdot
\bm{s}_0) \bm{\Omega} \tau_c^2}{1+(\Omega
\tau_c)^2},
\label{eq:overline-s}
\end{equation}
where $\bm{s}_0$ is the average value of the
initial spin. For an excitation with a periodic train of short pulses 
$W(t)\approx\delta (t-t_0-\tau_c)$,
and $\overline{\bm{s}}$
 is given by Eq.~(\ref{eq:soft}), by substituting
$\cos(\Omega (t-t_0))\rightarrow
\sin(\Omega \tau_c)/(\Omega\tau_c)$ and
$\sin(\Omega (t-t_0))\rightarrow
(1-\cos(\Omega\tau_c))/(\Omega
\tau_c)$. For our goal, the difference between the CW
and the pulsed excitation regimes is only quantitative. In this paper we
analyze (and present model calculations) only for the CW excitation regime.

In the short correlation time approximation 
$\bar{\bm{s}}\approx \bm{s}_{0}$, whereas in the opposite limit 
($\Omega \tau_c\ge 1$), the mean value of the electron
spin depends on the angular distribution of frequencies, 
$\bm{\Omega}$. For a random distribution (unpolarized nuclear
system)  $\langle \bm{\Omega} \rangle =0$, $(\bm{\Omega} \cdot
\bm{s}_0)\cdot \bm{\Omega} =\Omega ^2\bm{s}_0/3$, and
 $\bar{\bm{s}}\approx \bm{s}_{0}/3$. For a polarized
nuclear spin system  $\langle \bm{\Omega} \rangle\gg\bm{\Omega}_{fluc}$, and the  mean
value of the electron spin depends on the   direction of $\langle \bm{\Omega}\rangle$. In the absence of external magnetic
fields the ENSS has only one distinguished direction that is
determined by the photo-electron polarization, $\bm{s}_0$. For  $\langle \bm{\Omega} \rangle$  along
 $\bm{s}_0$  $(\bm{\Omega} \cdot \bm{s}_0)\approx \Omega s_0$,  and 
$\bar{\bm{s}}\approx \bm{s}_0$.

\subsection{Nuclear spin precession in the electron hyperfine field}\label{subsec:precession-hyperfine}

Let us now describe the mechanism of the nuclear polarization by the
resident electron.
We consider a slow nuclear spin precession that obeys the equation
\begin{eqnarray}
\frac{d\bm{I}_n}{dt}&=&\omega_n\left[\bm{s}(t)\times
\bm{I}_n\right]\nonumber\\
&\approx& \omega_n\left[\bar{\bm{s}}\times
\bm{I}_n\right]+\omega_n\overline{\left[\frac{d\bm{s}}{dt}\times
\bm{I}_n\right]}\delta t\nonumber\\
& &+\omega_n\overline{\left[\bm{s}\times
\frac{d\bm{I}_n}{dt}\right]}\delta t.
\label{eq:didt}
\end{eqnarray}
The average in this equation is done on the time region $\Delta t$, such that $\Omega\ll\Delta t\ll \omega$. 
The first term in the right-hand side of Eq.~(\ref{eq:didt}) gives the regular part of the nuclear
spin precession in the mean electron hyperfine field, $\bm{B}_{K,n}=-\omega_n\bm{\overline{s}}$, known as the 
 Knight field\footnote{In this paper the field is given in energy units. 
 The negative sign of the Knight field implies that the hyperfine energy has a minimum when $\bm{I}$ and $\bm{s}_0$ are parallel.}. 
The second term describes the dynamical polarization of the nuclei. For small nuclear polarization, $\langle I \rangle \ll I$, and:  
\begin{equation}
{\frac{\partial \bm{I}_n}{\partial
t}}|_{dp}=-\omega_n\tau_c\left\langle
\frac{\left[\left[\bm{\Omega} \times \bm{s}_0\right]\times
I_n\right]}{1+(\Omega\tau_c)^2}\right\rangle_I\approx
\frac{\omega_n^2\|I\|^2\bm{s}_0}{\langle
\omega^2\rangle \|s\|^2T_{1e}},
\label{eq:indp}
\end{equation}
where
\begin{equation}
T_{1e}(\Omega)=\frac{\Omega }{\langle
\omega^2\rangle}\left(2\|s\|^2\frac{(\Omega
\tau_c)}{3(1+(\Omega
\tau_c)^2)}\right)^{-1}
\label{eq:t1eomega}
\end{equation}
is the characteristic time of the longitudinal nuclear spin
relaxation, and  $\langle \omega^2\rangle
=\sum_n\omega_n^2/N$.
The third term\footnote{We neglect the part of Eq.~(\ref{eq:didt})
proportional to the square of the mean electron spin.} in Eq.~(\ref{eq:didt}),

\begin{equation}
\frac{\partial \bm{I}_n}{\partial
t}|_{I(t)}=\omega_n\sum_m
\omega_m\overline{\left[\bm{s}(t)\times
\int_0^t\left[\bm{s}(t_1)\times
\bm{I}_n\right]dt_1\right]},
\label{eq:didtit}
\end{equation}
represents two processes: relaxation of nuclear polarization (averaged
over initial electron spin direction),

\begin{equation}
\frac{\partial \bm{I}_n}{\partial
t}|_{rel}=-\frac{\omega_n^2\bm{I}_n}{\langle
\omega^2\rangle T_{1e}(\Omega,\tau_c)}(I_{n\|}+\frac{2+(\Omega\tau_c)^2}{2}I_{n\perp}),
\label{eq:inrel}
\end{equation}
and  nuclear spin precession,  $\partial\bm{I}_n/\partial
t|_{ind}=\left[\bm{\eta}_n\times\bm{I}_n\right]$, in the indirect hyperfine field given by

\begin{equation}
\bm{\eta}_n=-\frac{\omega_n^2\left[\bm{\Omega}
\times \bm{I}_n\right]\tau_c}{\langle
\omega^2\rangle T_{1e}(\Omega,\tau_c )}.
\label{eq:etan}
\end{equation}

\begin{figure}
\centering{
\includegraphics[width=8cm,clip]{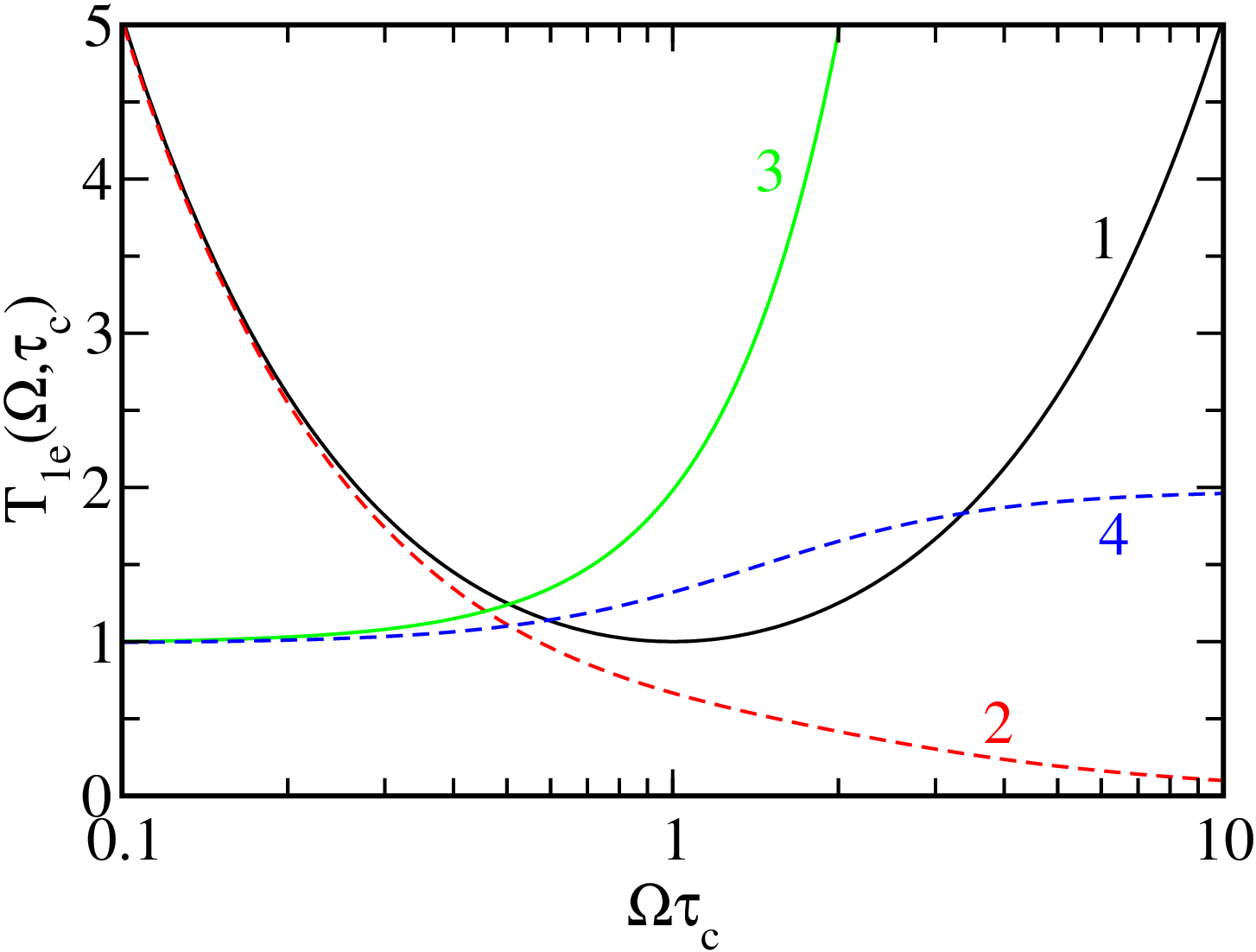}
}
\caption{\label{fig:3}
Nuclear spin relaxation time on the resident electron {\it vs.}~correlation time $\tau_c$  (curves 1,2),
 and {\it vs.}~frequency $\Omega$ (curves 3,4) of the electron spin precession in the nuclear hyperfine field.\\
Solid curves 1,3 show the relaxation time for the nuclear polarization component along $\bm{\Omega}$. 
Dashed curves 2,3 show the relaxation time of the nuclear polarization component transversal to $\bm{\Omega}$. 
Dependences of relaxation time on $\tau_c$  (1,2) were calculated for a constant value of $\Omega$. 
They are normalized to the value of $T_{\rm 1,e}$ at $\Omega\tau_c=1$. 
Dependences of relaxation time on $\Omega$ (3.4) were calculated for a constant value of  $\tau_c$, and are normalized 
to the value of $T_{\rm 1,e}$ at $\Omega=0$
The nuclear dynamic polarization is ineffective for $\Omega\tau_c\gg1$ because in this region the value of the relaxation time 
for the nuclear spin longitudinal component increases  fast, and the leakage factor in the Eq.~(\ref{eq:beta}) decreases. The difference between times $T_{\rm 1,e}$
 for longitudinal and transverse component of nuclear polarization in the region  $\Omega\tau_c\gg1$  increases 
 $\varepsilon$ in Eq.~(\ref{eq:beta}), but decreases nuclear polarization.
}
\end{figure}

(For a derivation of Eqs.~(\ref{eq:indp}), (\ref{eq:inrel}) and (\ref{eq:etan}) see the appendix.) Both
of these processes, \ie, relaxation of nuclear polarization and nuclear spin precession, are
determined by fluctuations of the electron hyperfine field. They are
proportional to  $\|s\|^2=3/4$  because
 $T_{1e}\propto \|s\|^{-2}$. 
The relaxation time versus $\Omega\tau_c$ is presented in Fig.~\ref{fig:3}.
 In the limit of
 short correlation time, all components of nuclear polarization relax
at the same rate\cite{re:meyer84}, $T_{1e}(0,\tau_c)^{-1}=2\|s\|^2\tau_c\cdot
\langle \omega^2\rangle /3$.

For intermediate
correlation times, the relaxation time for the component of $I$, $I_{\|}$, longitudinal to
$\bm{\Omega}$, increases as both $\Omega$ and $\tau_c$ increase, yielding\cite{re:abragam96} 
$T_{1e}(\Omega,\tau_c)=(1+(\Omega\tau_c)^2)T_{1e}(0,\tau_c)$ (solid curves 1,3 in Fig.~\ref{fig:3})
On the other hand, the relaxation rate of the polarization component, $I_{\perp}$, 
transverse 
to $\bm{\Omega}$,  behaves different depending on whether $\Omega$ is increase or
$\tau_c$ is increased. For $\tau_c$ increasing, $T_{1e,\perp}$ decreases monotonically (curve 2).
As $\Omega$ increases, $T_{1e,\perp}$ increases  
saturating at  $T_{1e\perp}(\infty,\tau_c)=2T_{1e}(0,\tau_c)$ (curve 4)\footnote{This saturation value is determined
by the fluctuation of the electron spin along  $\bm{\Omega}$. 
The average values of nuclear spins,
$\langle\bm{I}\rangle $,  and frequency, 
$\langle \bm{\Omega} \rangle $, are parallel to each other, and the 
relaxation rate for the polarized nuclei,
$I_{\|}\gg\|I\|/\sqrt{N}$,
 decreases as $(1+(\Omega\tau_c)^2)^{-1}$. For  states with low
polarization the fluctuation of the nuclear spin directions are more
important than the fluctuation of the modulus. For these states, increasing the correlation time decreases
relaxation rate by a factor of 2.}.

The nuclear polarization, its relaxation rate, and the indirect
hyperfine field, are all proportional to 
$\omega_n^2\tau_c\propto\|\psi(R_n)\|^4$,  and all have a
strong spatial dependence. They have a maximum  in the 
center of the QD, and they decrease towards the border.
In the limit of frozen nuclear field, \ie, short and intermediate
correlation times, $\omega_n\tau_c$ in Eqs.~(\ref{eq:indp}), (\ref{eq:inrel}) and (\ref{eq:etan}) 
is much smaller than 1. In Eq.~(\ref{eq:etan}), for an intermediate time, this parameter may be
rewritten as  $\omega_n/\Omega$, which is much less than 1, implying that
characteristic rates for nuclear spin polarization, relaxation, and
indirect hyperfine interaction are  many times less then $\omega_n$. 
These slow processes are
relevant only if they change the 
system's behavior, such as its nuclear polarization, Eq.~(\ref{eq:indp}), 
or its relaxation rate, Eq.~(\ref{eq:didtit}). 

By comparing the indirect field and the Knight field one can see 
that the former is significant only for  $s_0\le(\omega_n/\Omega)\cdot (\Omega\tau_c)^2/
(1+(\Omega\tau_c)^2)\le N^{-1/2}\ll1$. In the
following context we suppose that the electron polarization is high enough,
which allows us to ignore the indirect hyperfine interaction when it comes
together with the Knight field. (The indirect hyperfine
interaction between nuclei plays an important role in the relaxation of the
electron polarization transversal to a strong external magnetic
field\cite{re:yao06,re:deng06,re:witzel06}). 

\begin{figure}
\centering{
\includegraphics[width=7cm]{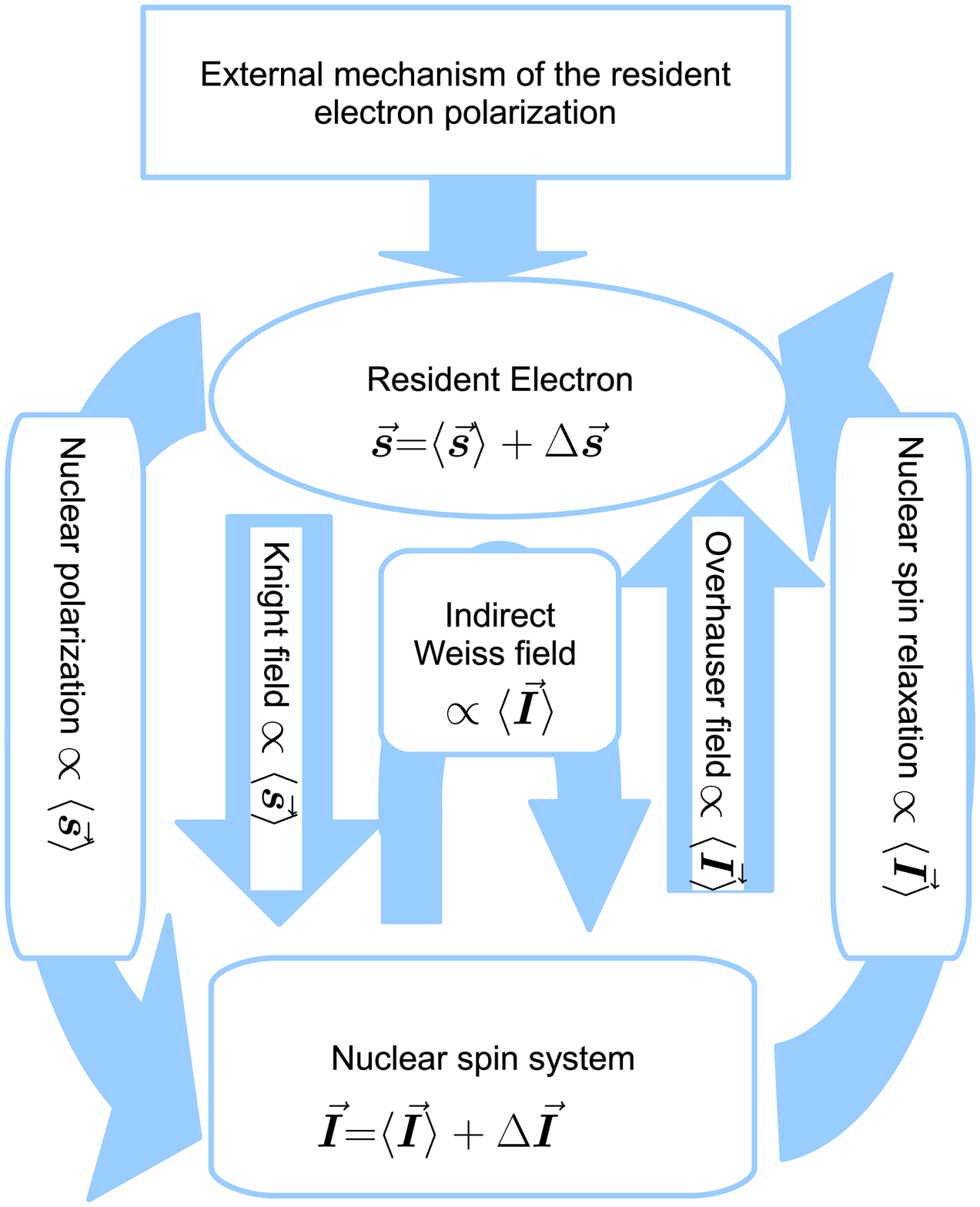}
}
\caption{\label{fig:2}
Different mechanisms for the hyperfine interaction
between the resident electron and the nuclei.\\
Nuclear spins precess in the mean Knight field, and are polarized by
the oriented resident electron. The rates of these effects are proportional to the average
electron spin. Polarized nuclei create a mean Owerhauser field on the
electron, and a mean indirect Weiss field on each other. Their polarization
also relaxes via interaction with the resident electron. The rates of these three mechanisms are
proportional to the value of nuclear polarization.}
\end{figure}

A sketch showing the main mechanisms of the hyperfine interaction's influence on the ENSS's 
behavior under sample illumination is presented in Fig.~\ref{fig:2}. This figure
shows three precessions (i) in the mean Knight field, created on the nuclei by a mean electron polarization, (ii) 
in the mean
Overhauser field, created on the electron by the mean nuclear spin, and (iii) in the mean indirect Weiss field, created on
the nuclei by the mean nuclear polarization.
It also contains two dissipation processes: nuclear polarization by the oriented electron 
through the fluctuation of nuclear spins, and nuclear polarization relaxation on fluctuations of
the electron hyperfine field.

When transitioning from short to intermediate correlation times, what changes most is the 
 dependence of the $T_{\rm 1e}$
time on $\Omega\tau_c$.
For $\Omega\tau_c\ll1$,
 the time 
$T_{1e}$ is proportional to $\langle \omega^2\rangle ^{-1}$, and also to $N^2$.
$T_{1e}$ is a decreasing function of $\tau_c$, and it reaches its minimum for $\Omega\tau_c=1$.
On the other hand, in the limit $\Omega\tau_c\gg1$, 
$T_{1e}(\Omega)\approx\tau_c(3\Omega^2)/(2\|s\|^2\langle\omega^2\rangle)$ is an increasing function of $\tau_c$. 
For a depolarized
nuclear system, \ie, when $\Omega \approx \Omega
_{fluc}$, the previous equation
gives  $T_{1e}\propto \tau_cN$, whereas
for a polarized nuclear system 
$T_{1e}\propto \tau_cN^2$, as we saw before. 

For a
typical GaAs QD (with $N=\text{10}^{5}$, 
$\langle \omega\rangle =\text{10}^{6}s^{-1}$, $\Omega_{\rm fluc}\approx 3\cdot\text{10}^{8}s^{-1}$), 
the shortest relaxation time for depolarized nuclei,
calculated from Eq.~(\ref{eq:t1eomega}) and condition 
$dT_{1e}/d\tau_c=0$, is  $T_{1e}\approx\text{10}^{-3}s$.
It is reached for 
$\tau_c^{(\rm min)}\approx 3\cdot\text{10}^{-9}s$.
On the other hand, for 100\% nuclear polarization,
 ${T_{1e}\approx\text{10}^{-1}s}$. This relaxation time is reached
when $\tau_c^{(\rm min)}=\text{10}^{-\text{11}}s$.
Since the short correlation time
approximation is valid for 
$\tau_c\ll\tau_c^{\left(\text{min}\right)}$, this
region  decreases by a factor of ${\sqrt{N}}$ for a highly polarized nuclear system.

The relaxation rate and precession frequency depend on the
nuclear position in the quantum dot. The balance between nuclear
dynamical polarization, Eq.~(\ref{eq:indp}), and relaxation, Eq.~(\ref{eq:inrel}), gives:
\begin{equation}
\langle \bm{I}_n\rangle=\frac{\|I\|^2\bm{s}_0}{\|s\|^2}.
\label{eq:insimple}
\end{equation}
Since the right-hand side of this equation does not contain $n$ or
$\tau_c$, the average nuclear
spin has the same value for all nuclei, \ie, $\langle
\bm{I}_n\rangle =\langle \bm{I}\rangle$. The previous statement is valid  both for short and
intermediate correlation times. 

Equation (\ref{eq:insimple}) is correct if we take into account only one
mechanism of nuclear spin relaxation, namely the hyperfine interaction with
fluctuations of electron polarization. Additional channels of
relaxation decrease the nuclear polarization. Often this
decrease can be described by introducing in Eq.~(\ref{eq:insimple}) the phenomenological
leakage factor\cite{re:meyer84}
$f=T_{1l}/{\left(T_{1e}+T_{1l}\right)\le 1}$. (Here
$T_{1l}$ is the relaxation time for additional
relaxation channels.) When $T_{1l}$ is finite, the 
leakage factor, ${f}$,  and the average polarization, $\langle \bm{I}\rangle$,
are monotonously decreasing functions of $T_{1e}$,
and have a maximum for $\Omega \tau_c\approx1$.

\subsection{Dipole-dipole interaction
between nuclear spins. Nuclear spin temperature}

The main additional channel of nuclear polarization relaxation is
determined by the dipole-dipole interaction between neighboring nuclear
spins. This interaction transfers nuclear
angular momentum to the crystal lattice with a characteristic time
$T_2\approx\text{10}^{-4}$s, which is much smaller than\cite{re:meyer84} $T_{1e}$.
 
As a result, the steady state value
of the quasi equilibrium nuclear polarization is $T_2/T_{1e}$ times less
than that predicted by Eq.~(\ref{eq:insimple}), and it has to be reached at a
time $T_2$.
Nevertheless, it is well known\cite{re:meyer84,re:dyakonov08}
from many experiments and theoretical calculations, that the optically
induced nuclear polarization cannot usually be ignored.
This polarization is  due to a decrease of the nuclear spin
temperature, $\Theta$. The effect
results from the balance of two energetic flows: cooling of
nuclear spins by oriented electrons in an external magnetic field 
$\bm{B}$: $J_{\rm cool}\propto -\left(\bm{B}\cdot \partial
\bm{I}/\partial t|_{\rm dp}\right)$, and heating of nuclear spins
by random fluctuations of electron polarization: $J_{heat}=\beta C$.
Here $\beta=(k_B\Theta)^{-1}$ is the inverse spin
temperature, and  $C={\rm dE}/d\beta$ is the
heat capacity of the nuclear spin system. In the short correlation time
approximation, $(\Omega\tau_c\ll1)$, and in a
spatially uniform external magnetic field\cite{re:meyer84}
\begin{equation}
\beta=f\frac{4I}{\mu_I}\frac{(\bm{B}\cdot \bm{s}_0)}{B^2+\varepsilon B_L^2},
\label{eq:beta}
\end{equation}
where $B_L^2$  is the
characteristic value of the random local field squared, and 
$\varepsilon$ is a number of the order of one.
$\beta$ has the opposite sign for
parallel and anti-parallel orientations of $\bm{B}$
and $\bm{s}_0$. (It is not surprising that the
spin temperature can be negative, since the energy of
the spin system is
limited both from above and below\cite{re:abragam96,re:landau80}). 
$\beta$ is positive if $\bm{s}_0$
is parallel to $\bm{B}$, in which case
 spin alignment decreases the nuclear spin energy.

In a general case, the field $B_L$ is a
result of dipole-dipole and indirect hyperfine interactions between
nuclei\cite{re:meyer84}. For the special case
of Eq.~(\ref{eq:etan}), the indirect interaction between two nuclei, $n$ and $m$, depends
on $n$ and $m$ only
through their product, $\omega_n\omega_m$,
and has no influence on the local field part of Eq.~(\ref{eq:beta}): $\varepsilon B_L^2\approx3B_{\rm dd}^2$. Here
 $B_{\rm dd}^2$  is
the square of the average dipole-dipole part of the local field. 

Eq.~(\ref{eq:beta}) is valid for high nuclear spin temperature, \ie,
for $\beta \mu_I\sqrt{B^2+\varepsilon B_L^2}\ll1$,  
and was derived for a spatially uniform $\bm{B}$
 and  $T_{\rm 1e}$.
An average Knight field should be included in the regular
external field, and the total magnetic field, $\bm{B}+\bm{B}_{K,n}$,
depends on the nuclear position inside the QD.

The approximation that ignores the spatial dependence of the
hyperfine interaction inside the dot is known\cite{re:ryabchenko83b,re:kozlov07} as the ``box model''.
For the ``box model'' the nuclear spin temperature and nuclear
polarization have the same value for all nuclei. In the real situation
of a spatially inhomogeneous hyperfine interaction, the nuclear spin
temperature and polarization cannot both be constant because
$\langle \bm{I}_n\rangle \propto \beta \bm{B}_{k,n}$. 

The dipole-dipole interaction between nuclear spins
produces an energy flow from the region with high spin
temperature to the
region with low spin temperature. It is commonly assumed that the
nuclear spin diffusion inside the area of electron localization is
suppressed by the strong gradient of the Knight field. (If the difference
in hyperfine splitting of the nearest nuclear spin levels is larger
than their dipole-dipole broadening, the flip-flop process between
nearest nuclear spins is suppressed by the energy conservation law.)
But one can show that the typical difference in the splitting of the
nearest nuclear spin levels for a spherical QD is about  $\hbar
\langle \omega\rangle s_0/N_R\approx \hbar \Omega
_{\rm max}s_0(4/I^3N^4)^{1/3}$, where $N_R\approx(N/4)^{1/3}$  is the number of nuclei 
along the QD radius. 
For a GaAs QD with $N=\text{10}^5$ 
this difference is about or less than  $\hbar\text{10}^4s^{-1}\approx \hbar\omega_{\rm dd}$,  and,
therefore,
there are no reasons for a strong suppression of
the spin diffusion. 

In the
limit of efficient spin
diffusion, the nuclear spin
temperature should have the same value for all nuclei, and
the equation for  $\beta$  contains
averaged values, \ie,
\begin{equation}
\beta=-\frac{4}{\hbar}f\frac{\langle\omega^3\rangle s_0^2}{\langle
\omega^4\rangle s_0^2+\varepsilon\langle
\omega_L^2\rangle \langle
\omega^2\rangle}.
\label{eq:betaaverage}
\end{equation}
Here  $\langle \omega^m\rangle=\underset{n}\sum \omega_n^m/N$  and
 $\langle \omega_L^2\rangle=\mu_I^2B_L^2/\hbar^2$.

In the limit of short correlation time, the difference
between the result of Eq.~(\ref{eq:betaaverage}) and the one obtained with the ``the box model'' is
only numerical. The dimensionless saturation value of the
spin temperature, $\tilde\beta\equiv \hbar \langle\omega\rangle\beta$,  is then 
$\langle\omega\rangle \langle\omega^3\rangle /\langle \omega^4\rangle$
 times less than that predicted by Eq.~(\ref{eq:beta}). It reaches
saturation if 
$s_0^2\gg\varepsilon\langle\omega_{\rm dd}^2\rangle \langle\omega^2\rangle /\langle \omega^4\rangle$. 
 (For a spherical quantum dot with infinite barrier
(SCDIB), we have: $\langle \omega\rangle^2/\langle \omega^2\rangle \approx0.36$, $\langle
\omega\rangle \langle \omega^3\rangle/\langle \omega^4\rangle \approx 0.22$, $\langle \omega^2\rangle^2/\langle \omega^4\rangle \approx0.17$,
and
$\varepsilon\langle\omega_{\rm dd}^2\rangle \langle\omega^2\rangle /\langle \omega^4\rangle\approx 5\cdot \text{10}^{-3}$).

In equilibrium, the mean value of the nuclear
spin in this field is
\begin{equation}
\langle \bm{I}_n\rangle=-\beta\frac{\|I\|^2}{3}(\hbar\omega_n)\bm{s}_0.
\end{equation}
When there is a single global nuclear spin temperature for all points in the QD, the maximum average
 nuclear polarization within the SCDIB model is  4 to 5 times smaller than the one predicted by the ``the box model''. 

In section~\ref{subsec:precession-hyperfine}, we saw that
$T_{1e}$  as function of 
$\tau_c$ has a minimum for 
$\tau_c=\Omega ^{-1}$. For $\tau_c>\Omega ^{-1}$,
increasing  the correlation time and the frequency 
$\bm{\Omega}$  increases $T_{1e}$. The
nuclear polarization increases $\bm{\Omega}$, implying that if one wants to optimize
the nuclear polarization by varying the
correlation time, one has to start from a unpolarized nuclear system
with $\Omega_{\rm fluc}\tau_c\ll1$. At the same time, $\tau_c$
should be long enough to achieve the condition 
$\Omega \tau_c=1$, and to  maximize the
leakage factor in the final polarized state. Then, the best regime for generating high
nuclear polarization is on the border between short and intermediate
correlation times.

\begin{figure}[h]
\centering{
\includegraphics[width=7cm,clip]{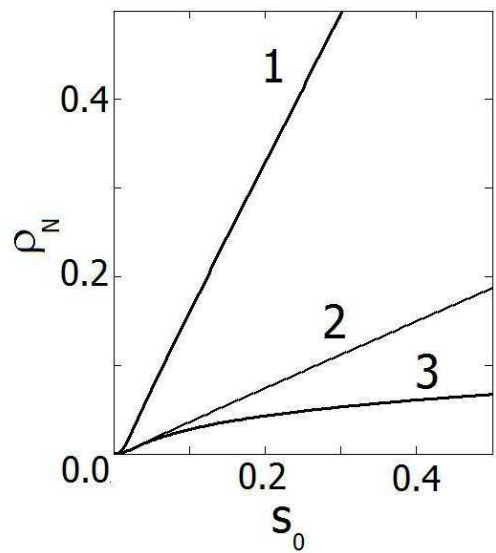}
}
\caption{\label{fig:3mx}
Dependence of nuclear polarization on the
mean spin of the resident electron. Here $T_{1e}(0)=T_{1l}$.
Curve 1 was calculated using the ``box model'' with
$\Omega_{\rm max}\tau_c\ll1$,
curve 2 using the SCIB model with $\Omega_{\rm max}\tau_c\ll1$, and
curve 3 using the SCIB model with $\Omega_{\rm max}\tau_c=\text{10}$.}
\end{figure}

For $\Omega\tau_c\gg1$, the anisotropy of the nuclear spin relaxation on the electron 
 also renormalizes the parameter
$\varepsilon\approx 3+(\Omega\tau_c)^2$. We can
neglect this effect for $\Omega \tau_c\le 1$.
In Fig.~\ref{fig:3mx}, results of the nuclear spin polarization
calculations for the ``box model'' (curve 1) and SQDIB model (curves 2,3)  are
presented. All calculations are done for 
$T_{1e}(0)=T_{1l}$. Curves 1 and 2 are calculated in the
short correlation time approximation, curve 3 for 
$\Omega_\text{max}\tau_c=10$.
One can see that the nuclear polarization for the SQDIB model is about 5 times
less than for the ``box model''.  For $\Omega_{\text{max}}\tau_c=10$, increasing the time 
$T_{1e}$ by increasing the nuclear polarization
additionally decreases the leakage factor and slows down an increase of the nuclear polarization.

\section{Electron-nuclear Spin System in the Limit of Long Correlation Time (in darkness)}\label{sec:darkness}

In this section we consider the relaxation of an isolated ENSS
in the limit of long correlation time. The nuclear polarization 
by spin-oriented carriers is extremely ineffective. 
We will consider only spin relaxation of an isolated ENSS in the darkness.
In section~\ref{subsec:conservation} we discuss
the main thermodynamic potentials that characterizes the isolated spin
system. We connect these potentials to the relaxation of ENSS parameters 
under illumination and discuss the relaxation mechanism. In 
section~\ref{subsec:dipole} we present the results of a numerical simulation of the QD
spin relaxation due to the dipole-dipole interaction that
transfers nuclear spin into the crystalline lattice angular momentum.
This process is controlled only by the state of the QD ENSS, and is independent of the QD's environment.

\subsection{Conservation laws and thermodynamic potentials of the ENSS}\label{subsec:conservation}
The $\tau_c$ of the ENSS largely increases
in the darkness. At 
liquid helium temperatures the characteristic time of spin relaxation
 for electrons on phonons is about seconds, whereas for nuclei it ranges 
 from days to years\cite{re:dyakonov08,re:abragam96,re:erlingsson01,re:khaetskii00}.
The direct transfer of spin angular momentum to the crystal by the
dipole-dipole interaction between nuclear spins is the main mechanism
for the relaxation of the spin polarization. This interaction is
also responsible for the energy diffusion from the quantum dot to the
neighboring nuclei\cite{re:paget82,re:makhonin08,re:nikolaenko09}.

In the zero-th order approximation, we keep only the hyperfine
interaction and switch off all other
interactions, which makes it easier to determine some integrals of motion.
The Fermi interaction conserves  energy, $E=\hbar \Omega s_\Omega$,
and total spin, $\bm{F}=\bm{I}_\Sigma+\bm{s}$. The total spin of
the nuclei,  $\bm{I}_\Sigma$, is many orders of
magnitude larger than that of the electron. Therefore, with high precision
$\bm{I}_\Sigma$ is also conserved. Moreover, as a result of
the adiabatic approximation, 
$\Omega \gg\omega_n$, and the electron spin projection,
$\bm{s}_\Omega$, along $\bm{\Omega}$  is also conserved and must be quantized, \ie,
$s_\Omega=\pm 1/2$. 
The  conservation of $\Omega$ 
follows from the conservation of energy and
$s_\Omega$.

Under illumination every nucleus is acted on by
an average Knight field, whose value and direction is determined by the
average electron spin, whereas in the darkness the ensemble of quantum dots
decomposes in two sub-ensembles (nuclear spin polarons) with defined value of electron spin
projection, $s_\Omega$, on the nuclear hyperfine field, \ie, $s_\Omega=\pm 1/2$,
 and energies $\pm \hbar \Omega /2$.
The probability to find a quantum dot in one of these sub-ensembles is
given by the conservation of  energy,  frequency, and total nuclear spin.

We will assume that the transition from illumination to
darkness is sharp, 
and that in the initial state the total nuclear polarization is larger
than its fluctuations, \ie, $\langle I\rangle\gg1/\sqrt{N}$.
Considering the integrals of motion of the system, we can write
the ENSS probability distribution $\Phi$, as a function of three thermodynamic potentials:
an electron spin potential, $\varsigma$, an inverse nuclear spin temperature, $\chi_{s_\Omega}$,
 for each sub-ensemble,
and a nuclear spin potential,
$\bm{\xi}(\chi_{s_{\Omega}})$, 
\begin{equation}
\Phi(\varsigma,\chi,\bm{\xi})\approx
\frac{\exp\left\{(\varsigma
s_\Omega)
-\left(\chi_{s_{\Omega}}\hbar \Omega s_\Omega\right)+\left(\bm{\xi}\cdot\bm{I}_\Sigma\right)\right\}}
{(4\pi)^N(\exp\{\varsigma/2\}+\exp\{-\varsigma/2\})}
\end{equation}

At the last moment of illumination, the total nuclear spin mean value, $\langle \bm{I}_\Sigma\rangle$,  and
the electron spin, $\bm{s}_0$, are directed along the Knight
field, and the nuclear inverse  spin temperature under illumination is $\beta$.
Therefore, after switching off the light, the nuclear inverse spin temperature in the darkness is
\begin{equation}
\chi_{s_{\Omega}}\approx
\beta\frac{s_0}{s_{\Omega }}.
\label{eq:chi}
\end{equation}
Moreover,
nuclear and electron spin potentials are given by
\begin{equation}
\bm{\xi}=\frac{3\langle \bm{I}_\Sigma\rangle
}{\langle \|\bm{I}_\Sigma\|^2\rangle},
\label{eq:xi}
\end{equation}
 and
\begin{equation}
\varsigma=\text{ln}\frac{1+2s_0}{1-2s_0}.
\label{eq:varsigma}
\end{equation}
In Eq.~(\ref{eq:chi}) we took into account that for  a cooled QD nuclear spin
system  $\langle \|\bm{I}_\Sigma\|^2\rangle
\approx \|\bm{I}\|^2N+\|\langle \bm{I}_\Sigma\rangle^2\|$. The first part of the right-hand side in this equation
describes the fluctuation of the total nuclear spin, and the second part
the square of the mean value of the total nuclear spin in the electron
hyperfine field. Equations~(\ref{eq:chi}), (\ref{eq:xi}),
and (\ref{eq:varsigma}) completely determine the initial
state of the system. From Eq.~(\ref{eq:chi}) one can see that
$\chi_{+1/2}=-\chi_{-1/2}$. In the following we will consider only the sub-ensemble with
a positive spin temperature, and omit $\chi$'s sub-index.

\begin{figure}
\centering{
\includegraphics[width=8cm,clip]{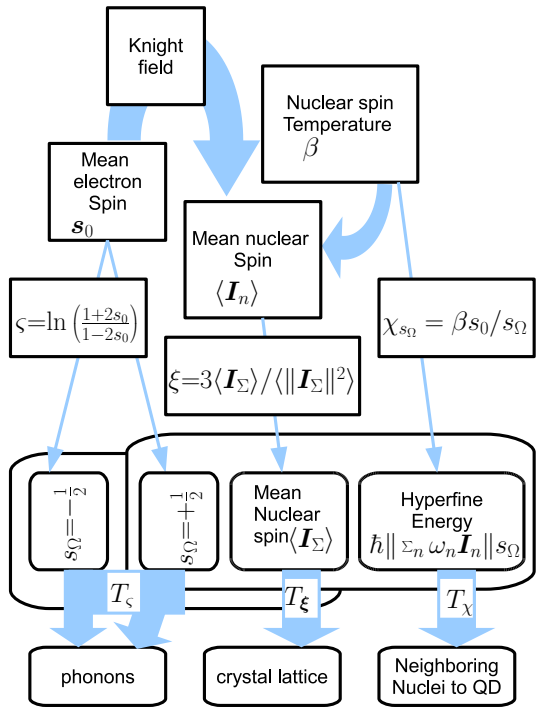}
}
\caption{\label{fig:4}
Relations between the ENSS's description  under
illumination and in the darkness. Under illumination the ENSS state
is described by the average electron spin, $s_0$, and the inverse nuclear spin temperature,
 $\beta$. The Knight field of the electron creates
the mean polarization of the cooled nuclei. After a sharp transition to
darkness, the ensemble of quantum dots splits in two sub-ensembles with two different
electron spin projections, $s_\Omega=\pm 1/2$, on the nuclear hyperfine field. 
The ENSS is
characterized by an electron potential, 
$\varsigma$, by a nuclear potential,
$\bm{\xi}$, and by a nuclear inverse 
spin temperature, $\chi_{s_{\Omega}}$. The relaxation of these thermodynamic
potentials is due to the dipole-dipole interaction between
nuclear spins and electron phonon interaction. 
This relaxation transfers nuclear spin to the crystalline lattice (on a time
$T_{\bm{\xi}}$),  energy diffusion from the quantum
dot to the environment (on a time $T_\chi$), and energy from the scattering of phonons (on a time 
$T_\varsigma$).}
\end{figure}

In the darkness, the relaxation of the potentials 
$\bm{\xi}$, $\chi$, and $\varsigma$, is due to the dipole-dipole interaction between
neighboring nuclei, and to phonon scattering. Each relaxation potential
has its own relaxation time,  for $\bm{\xi}$,  
$T_{\bm{\xi}}$; for $\chi$, $T_{\chi}$; and for $\varsigma$, $T_\varsigma$. 
The
dipole-dipole interaction does not conserve total spin, $\bm{I}_\Sigma$. As a result, the angular
distribution of $\bm{I}_\Sigma$ tends to the isotropic distribution, and
$\bm{\xi}$ tends to zero. 
But the dipole-dipole interaction conserves total energy, and the modulus of $\bm{I}_\Sigma$ fluctuates
around its average value $\langle I_\Sigma\rangle(\chi)=\chi\hbar\langle\omega\rangle N \|I\|^2/6$.
In 
section \ref{subsec:dipole} we demonstrate that the rate at which $\bm{\xi}$
relaxes depends on the value of 
$\bf{I}_\Sigma$, or, in other words, that $T_{\bm{\xi}}$  is a function of
$\chi$.

The relaxation of $\bm{I}_\Sigma$,
and of the frequency, $\langle \Omega \rangle$, is a result of the
nuclear spin energy flow out of the quantum dots provided by
spin-spin interactions between nearest-neighboring nuclei.
As a result of this process, $\chi$ tends to 0, and the frequency of electron spin precession relaxes
from its initial value to an asymptotic fluctuation value given by
$\Omega _{\rm fluc}=\sqrt{\|I\|^2N\langle\omega^2\rangle}$, which is non zero for a
finite system. The rate of this relaxation process depends on the
value and sign of the nuclear spin temperature around the QD.

The state with $\chi\approx0$ and $\varsigma\neq 0$ 
can be identified as a
fluctuating nuclear spin polaron{\footnote{A fluctuating magnetic polaron was
investigated in the Raman spectroscopy of  diluted magnetic
semiconductors in
Refs.~\onlinecite{re:alov81,re:dietl82,re:ryabchenko83b,re:heiman83}.}.
In  the fluctuation polaron state the electron ``remembers'' its spin direction
 after total nuclear spin relaxation takes place. To put it another way, the potential $\varsigma\neq 0$ describes not electron polarization
but correlation between electron spin and nuclear field. The
relaxation of $\varsigma$  is caused by  the flipping of the electron spin.
 The energy of this transition, $\hbar \Omega $, cannot be taken from the nuclear
spin system as $\Omega\gg\omega_L$,
but is due to
phonon-scattering instead\cite{re:abragam96,re:erlingsson01,re:khaetskii00}.
A sketch of connections between ENSS parameters and their
relaxations is presented in Fig.~\ref{fig:4}.

In short, the relaxation of $\varsigma$
 and $\chi$
 is due to the open character of the QD ENSS, \ie, it
depends on the environment. We will not discuss it later.

\subsection{The dipole-dipole relaxation of an isolated quantum
dot}\label{subsec:dipole}

The dipole-dipole relaxation of the nuclear spin system is
commonly\cite{re:abragam96} characterized by a time 
$T_2\propto \sqrt{\langle\omega_{\rm dd}^2\rangle}$. For GaAs\cite{re:meyer84}
${T_2\approx\text{10}^{-4}s}$. The estimations
in Refs.~\onlinecite{re:merkulov98,re:oulton07,re:dyakonov08}
demonstrate that, for a QD composed 
of an electron and polarized nuclei (nuclear spin polaron), the relaxation
time of spin polarization should increase as
\begin{equation}
T_{\bm{\xi}}(N,\rho_\Omega)\approx T_2N\rho_\Omega^2,
\label{eq:tnrho}
\end{equation}
 where
$\rho_\Omega^2=\langle I_\Omega^2\rangle /N^2\|I\|^2$ is determined
by the nuclear spin temperature, $\chi$.
$I_\Omega$ is the projection of the total
nuclear spin on the direction of the nuclear hyperfine field. 

We now compare the estimation of Eq.~(\ref{eq:tnrho}) with the results of
our numerical simulations of the ENSS spin dynamics.
For this we calculate the nuclear spin correlator, 
\begin{equation}
G(t,\rho_\Omega)=\frac{\int \bm{I}_\Sigma(t')\bm{I}_\Sigma(t'+t)\cdot\text{dt}}{\int
\bm{I}_\Sigma(t')\cdot\bm{I}_\Sigma(t')\text{dt}}.
\label{eq:correlator}
\end{equation}
Our numerical model considers a spherical quantum dot containing $N_{\rm mod}$ nuclei
located on a cubic crystal lattice.
To make the calculation feasible we take $N_{\rm mod}$ to be of the order of  hundreds,
and, therefore, much less than the number of 
 nuclei in a real quantum dot. However, $N_{\rm mod}\gg1$, and the adiabatic approximation may be used
 to model the system. Therefore, the electron spin has a constant projection, $s_\Omega$, on the
total nuclear field given by $s_\Omega=\pm 1/2$, whereas the nuclear spin precesses around the
hyperfine electron field (directed along $\bm{\Omega} =\sum_n\omega_n\bm{I}_n$) with frequency
$\omega_n/2$.

As the frequency unit we take the mean frequency of nuclear
spin precession, $\langle\omega\rangle/2\equiv 1$. As the length unit
we take the distance between nearest nuclei, $r_0\equiv 1$.
We use the following equation to calculate the precession of the nuclear spin 
at site $n$ in the magnetic field created by neighboring spins,

\begin{equation}
\frac{\partial \bm{I}_n}{\partial
t}|_{\rm dd}=\gamma\sum_{m\neq n}\frac{\left[
\left(\frac{1}{3}\bm{I}_{m}-\bm{e}(\bm{e}\bm{I}_m)\right)\times \bm{I}_n\right]}{d_{nm}^3}.
\end{equation}

Here $\bm{e}$ is the unit vector joining
sites $n$ with spin $m$,
$d_{nm}$
is the distance between these sites,
$\gamma=\omega_D/\langle\omega\rangle \approx 5.7\times\text{10}^{-2}$, and
$\omega_D=\mu_n^2/(\hbar r_0^3)$ is the characteristic frequency of the
nuclear spin precession in the  field of nearest neighbor dipoles.

The sign of the electron spin affects only the direction of
the nuclear hyperfine precession, but has no influence on the dipole
relaxation. For this reason, we simulated only the sub-ensemble with positive
spin temperature. The initial distribution of nuclear spins  is determined by the inverse temperature, $\chi$, with
natural units 
$\hbar \langle \omega\rangle $. This initial spin
distribution was generated by a random process using the Boltzmann
distribution function. In order to decrease the effects of crystal magnetic
anisotropy in the initial
state\footnote{As is
well known from magnetism theory, the energy of magnetic anisotropy
for a cubic spin lattice is proportional to 
$I_{\Sigma,x}^4+I_{\Sigma,y}^4+I_{\Sigma,z}^4+O(I_\Sigma^6)$. It is important only for high enough nuclear polarization},
$\Omega$ was directed along the [111] axis.

The calculated dependence of  $\rho_\Omega$  on the inverse spin temperature, $\chi$,
 is presented in Fig.~\ref{fig:5}. The numerical results  are in a good agreement with the simple  theoretical equation,
 \begin{equation}
\langle \rho_\Omega\rangle \approx
\sqrt{\langle L\rangle ^2+\langle \mathit\omega\rangle^2/(N\langle \omega^2\rangle )}.
\label{eq:rhoOfOmega}
\end{equation}
Here $\langle L\rangle $ is the Langevin
function, averaged on the QD volume, and is given by
\begin{equation}
\langle L(x)\rangle =3\int_0^\pi\left[
\frac{e^{x\cdot\frac{\sin^2(r)}{r^2}}+e^{-x\cdot
\frac{\sin^2(r)}{r^2}}}{e^{x\cdot
\frac{\sin^2(r)}{r^2}}-e^{-x\cdot
\frac{\sin^2(r)}{r^2}}}-\frac{r^2}{x\cdot
\sin^2(r)}\right]r^2\text{dr},
\label{eq:lofx}
\end{equation}
where $x=\chi\omega_0$, and the index 0 indicates the nucleus at the center of the QD.
The second term under the square root in Eq.~(\ref{eq:rhoOfOmega})  describes the
nuclear spin fluctuation, which is important in the limit of extremely high temperature, when
$\chi\hbar \langle\omega\rangle \le 1/\sqrt{N}$.

To control the numerical precision, we checked the
conservation law for the energy and the total nuclear spin of the
system without dipole-dipole interaction (see Fig.~\ref{fig:6}(a)). After
introducing the dipole-dipole interaction (Fig.~\ref{fig:6}(b)) the total
energy is still conserved (curve 3), but now all three components of
the total spin have a random behavior (see, for example, curve 1 for
$I_{\Sigma,x}(t)$). The total
nuclear spin, $\|I_\Sigma(t)\|$,  fluctuates
around its mean nonzero value that is related to the total hyperfine energy,
 $\langle \omega\rangle\langle I\rangle \approx \langle \Omega \rangle \propto E$.
The fluctuations of hyperfine energy are negligibly small because they
reflect very weak transfers of energy from hyperfine to dipole-dipole
reservoirs and vice versa.
\begin{figure}
\centering{
\includegraphics[width=7cm,clip]{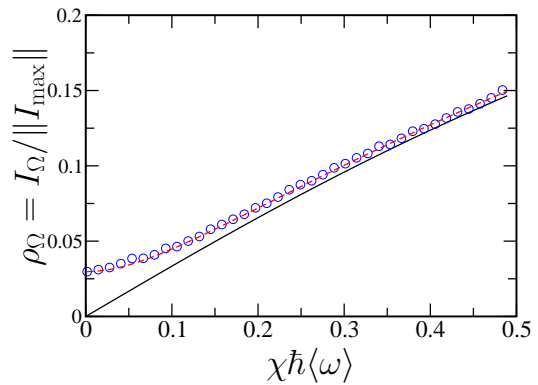}
}
\caption{\label{fig:5} Nuclear spin polarization as a function of the inverse
nuclear spin temperature for a spherical quantum dot with infinite
wall. The solid curve  is the averaged Langevin distribution, Eq.~(\ref{eq:lofx}), for
infinite numbers of nuclei, the dashed curve, the calculation for the SQDIB model, 
Eq.~(\ref{eq:rhoOfOmega}),  with $N=489$. Circles are the Monte Carlo simulation, also with $N=489$.}
\end{figure}

In Fig.~\ref{fig:7} is the calculated correlator $G(t)$, Eq.~(\ref{eq:correlator}), for various
 nuclear spin temperatures, as indicated. One can see
that an increase of the nuclear polarization, from curve 4 with
$\rho_\Omega=0.04$,
to curve 1 with $\rho_\Omega=0.152$, decreases  the spin relaxation rate. On a
short time scale, the time dependence of the spin correlator can be
approximated by a Gaussian distribution:
\begin{equation}
G(t,\langle I\rangle,N)\approx
\exp\left\{-\frac{t^2}{T_{\bm{\xi}}^2(\beta,N)}\right\}.
\label{eq:gaussian}
\end{equation}

\begin{figure}
\centering{
\includegraphics[width=7cm]{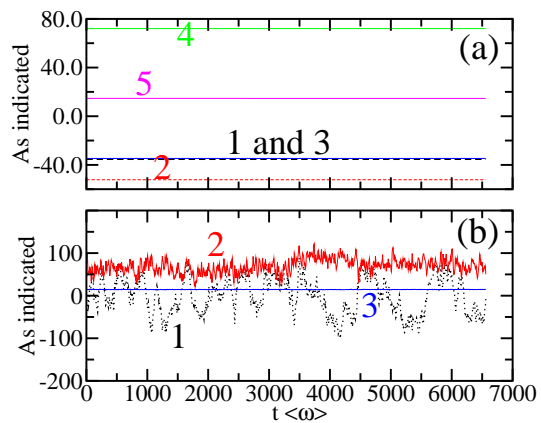}
}
\caption{\label{fig:6} Conservation of the quantum dot's
parameters. (a) With the dipole-dipole interaction switched off, the spin components
$I_{\Sigma,x}$, $I_{\Sigma,y}$, and $I_{\Sigma,z}$, given by curves 1, 2, and 3, respectively,
the spin modulus, $I_\Sigma$, given by curve 4, and
the energy of
the system, $E$, given by curve 5, are all conserved.
(b) 
When the
dipole-dipole
interaction is switched on, the spin components are no longer conserved,
and change their values chaotically instead (curves 1, 2, $z$ component is now shown).
But the spin modulus, curve 3, is still conserved within numerical error.
}
\end{figure}

\begin{figure}
\centering{
\includegraphics[width=7cm,clip]{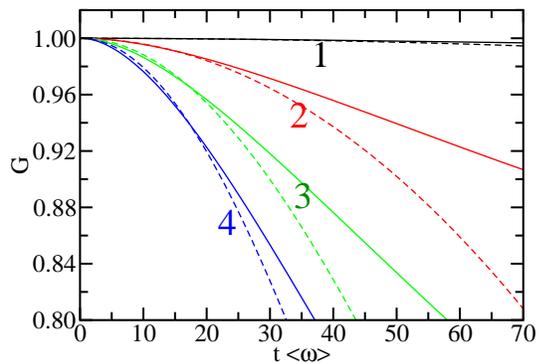}
}
\caption{\label{fig:7}
Spin correlator, $G$, {\it vs.}~time 
for different  nuclear temperatures, as indicated,
on a quantum dot with $N=1365$ nuclei. Solid lines are for the numerical simulation,
whereas dashed lines are a Gauss approximation with fitting parameter 
$T_{\bm{\xi}}$ following Eq.~(\ref{eq:gaussian}).
(1)  $\beta\hbar \langle\omega\rangle \approx 3.5$,
$\rho_\Omega=0.410$, $T_{\bm{\xi}}\langle\omega\rangle =820\pm15$,
(2) $\beta\hbar \langle\omega\rangle \approx 0.18$($\rho_\Omega \approx0.051$,
$T_{\bm{\xi}}\langle \omega\rangle =160\pm15$,
(3) $\beta\hbar \langle\omega\rangle \approx 0.078$, $\rho_\Omega=0.036$, 
$T_{\bm{\xi}}\langle \omega\rangle =90\pm15$, 
(4) $\beta\hbar \langle\omega\rangle \approx 0.05$
$\rho_\Omega=0.027$, $T_{\bm{\xi}}\langle \omega\rangle =60\pm15$.}
\end{figure}

This equation allows us to quantitatively compare the result of
our numerical experiment with the estimation of
Eq.~(\ref{eq:tnrho}). Figure~\ref{fig:8} shows our numerically calculated
value for $T_{\bm{\xi}}(\beta,N)$ for 4
 different spherical quantum dots, with $N$ equal to 251, 485, 895
and 1365 nuclear spins.  There
$T_{\bm{\xi}}(\beta,N)\langle\omega\rangle/N$ {\it vs.}\,$\rho_\Omega^2(\beta)$ is presented; 
the polarization relaxation rate decreases fast with increasing nuclear
polarization. 
$T_{\bm{\xi}}(\beta,N)\langle \omega\rangle/N$  has
approximately the same value for different N, and calculated points
for different system sizes are grouped around a straight line. As follows
from Eq.~(\ref{eq:tnrho}), the slope of this line gives the characteristic time, which in this case is
 $T_2\langle \omega\rangle \approx 60$ or $T_2\gamma\approx 3$.
These results are also in good agreement with Eq.~(\ref{eq:tnrho}); they 
demonstrate the universal character of the predicted dependence of
the dipole-dipole relaxation time for the nuclear spin potential 
$\bm{\xi}$ on $\rho_\Omega(\beta)$ and system size, $N$. 
\begin{figure}
\centering{
\includegraphics[width=7cm,clip]{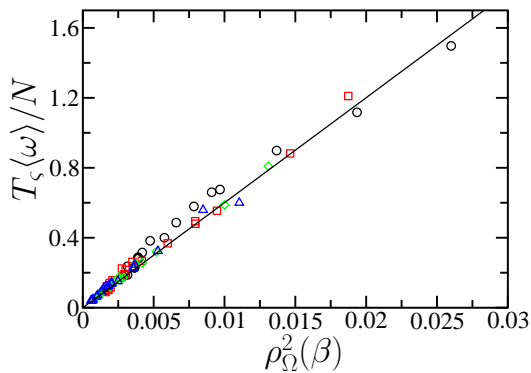}
}
\caption{\label{fig:8}
Dependence of nuclear spin relaxation time on
nuclear spin polarization.
Circles are for $N=251$, squares
for $N=485$, diamonds  for $N=895$,  and triangles
for $N=1365$.
 All points coincide well
with the straight line calculated from Eq.~(\ref{eq:correlator}), with slope 
$T_2\gamma\approx 3$.
}
\end{figure}

\section{Summary}\label{sec:summary}

{\it (1)  We demonstrated that for the short
correlation time limit the rate of nuclear relaxation on electrons is
proportional to the correlation time, \ie, $T_{1e}\propto\tau_c^{-1}$. 
In the opposite limit of long
correlation time ($\Omega\tau_c\gg1$) we found that $T_{1e}\propto\tau_c$.} The maximal rate of nuclear
polarization by the QD's electron is reached for an intermediate value of 
correlation time, $\Omega \tau_c\approx 1$.
In this case the leakage factor reaches its maximum. Nuclear
polarization in the long correlation time regime is not efficient, which 
 follows directly from the general equations in Ref.~\onlinecite{re:dyakonov75},
that connect nuclear polarization and relaxation rates with
electron spin correlator. 

{\it (2) 
The nuclear spin diffusion inside a QD increases the average nuclear
spin temperature and decreases nuclear polarization.} The diffusion is a
result of the spatial dependence of the spin relaxation rate and of the Knight field. Nuclear polarization diffuses from the QD's center to its
periphery. In the vicinity of the barrier, the Knight field is comparable with the local dipole
field, and strong dipole-dipole relaxation destroys nuclear
polarization. This effect decreases the mean value of the nuclear
polarization by a factor of more than 4. One  should  take this effect into account when 
experimentally describing  the value of the nuclear polarization.

{\it (3) The indirect hyperfine field is contributed by
a macroscopically high number of QD nuclei. The strength of the indirect hyperfine interaction between nuclei increases
for longer $\tau_c$, and reaches a maximum for $\Omega\tau_c\ge 1$.}
 This maximum  is about $\hbar\omega_n^2/\Omega$. It is inversely
proportional to nuclear polarization. Usually the field is less than
the Knight field, $\hbar\omega_ns_0$. The indirect field plays an important role in the problem of
electron spin dephasing\cite{re:yao06,re:deng06,re:witzel06}. It may also be important in the
realization of dynamic nuclear
self-polarization\cite{re:dyakonov72}, where electron polarization and Knight field are equal to
zero.

{\it (4) In the
regime of long correlation time, the state of the quantum dot 
 is characterized by three thermodynamic
potentials: $\varsigma$, $\bm{\xi}$ and $\chi$.}
These potentials directly affect the average electron spin, the average nuclear spin, and the 
nuclear spin temperature, respectively. The relaxation of the nuclear
spin potential $\bm{\xi}$  by 
the dipole-dipole
interaction affects the transfer of angular momentum to the crystal lattice. The diffusion of energy to the QD's
environment is the main mechanism for the relaxation of the inverse nuclear spin
temperature, $\chi$. The
relaxation of the electron spin potential 
$\varsigma$ is connected with phonon scattering.

{\it (5) Our numerical simulation showed the ENSS's
behavior for the simple case of a pure hyperfine interaction between
a resident electron and nuclei, and for the real case of an additional dipole-dipole interaction between nearest nuclei.} 
They
demonstrated a suppression of the $\bm{\xi}$ relaxation that was caused by a decrease 
in the nuclear spin temperature. The
dipole-dipole relaxation time was found to be proportional to the number of nuclei, and to the nuclear polarization squared. These results are
in  good agreement with the analytical expression introduced in
Refs.~\onlinecite{re:merkulov98,re:oulton07}.

\begin{acknowledgments}
The authors are thankful to M.I. Dyakonov, V.G. Fleisher,
and S.M. Ryabchenko for fruitful discussions.  
I.A.M. thanks the Program of Russian Academic of Science, Spin Phenomena in Semiconductor Nanostructures and Spintronics.
We acknowledge the support of
the Center for Nanophase Materials Sciences, sponsored by the Scientific User
Facilities Division, Basic Energy Sciences, U.S. Department of Energy, under
contract with UT-Battelle.
\end{acknowledgments}

\appendix*
\section{Derivation of Nuclear Spin Precession Equations}

To derive Eq.~(\ref{eq:indp}) we introduce Eq.~(\ref{eq:indp}) in Eq.~(\ref{eq:didt}) 

\begin{equation}
\overline{\frac{d\bm{s}}{\rm dt}\delta t}=\bm{s}_0-\bar{\bm{s}}=
\frac{(\Omega\tau_c)^2\bm{s}_0-\left[\bm{\Omega} \times
\bm{s}_0\right]\cdot\tau_c-(\bm{\Omega} \cdot \bm{s}_0)\cdot \bm{\Omega} 
\tau_c^2}{1+(\Omega\tau_c)^2}
\label{eq:dsdtdeltat}
\end{equation}
For small nuclear polarization $|\langle \bm{I}\rangle|\ll\|I\|$, and  the vector product
in the right-hand side of Eq.~(\ref{eq:dsdtdeltat}) plays the main role in the average rate of nuclear
polarization, \ie,
$\langle (\Omega\tau_c)^2\left[\bm{s}_0\times
\bm{I}_n\right]\rangle_I$, where $\langle (\Omega \bm{s}_0)\left[\bm{\Omega} \times
\bm{I}_n\right]\tau_c^2\rangle_I\ll\langle \left[\left[\bm{\Omega} \times
\bm{s}_0\right]\times \bm{I}_n\right]\rangle_I\tau_c\approx-\frac23\omega_n\tau_c\|I\|^2s_0$.
Here $\langle \cdots\rangle_I$ represents an average over the nuclear spin
direction. Eq.~(\ref{eq:indp}) directly follows from this.

It follows from (Eq.~\ref{eq:didtit}) that 
\begin{eqnarray}
& \left\langle
 \frac{d\bm{I}_n}{\rm dt}\right\rangle_s  = 
\frac{\omega_n^2}{\tau_c}\left\langle\int_0^\infty
\left\{(\bm{s}(t)\cdot \bm{I}_n) \int_0^t \bm{s}(\tau) d\tau\right.\right.-\nonumber\\
& - \left.\left.\bm{I}_n \left(\bm{s}(t)\cdot
\int_0^t \bm{s}(\tau)d\tau\right)\right\}\exp\left\{-t/\tau_c\right\}{\rm dt}\right\rangle_s,
\label{eq:didtaverage}
\end{eqnarray}
where $\langle \cdots\rangle_s$ represents the average over the initial spin
direction. The time dependence of electron spin is given by Eq.~(\ref{eq:soft}). Eq.~(\ref{eq:didtaverage})
is an odd function of the electron spin, and to linear approximation it
does not depend on polarization. For a random initial electron spin
distribution $\langle s_\alpha(0)s_\beta(0)\rangle=\delta_{\alpha,\beta}\|s\|/3=\delta_{\alpha,\beta}/4$,
yielding
\begin{eqnarray}
& & \frac{1}{\tau_c}\left\langle \int_0^\infty\left(\bm{s}(t)\cdot
\int_0^t\bm{s}(\tau)d\tau\right)\exp\left\{-\frac{t}{\tau_c}\right\}d\tau_c\right\rangle_s  =  \nonumber\\
& & \frac{\tau_c}{4}\left(1+\frac{2}{1+(\Omega\tau_c)^2}\right),
\end{eqnarray}
and
\begin{eqnarray}
& & \frac{1}{\tau_c}\left\langle \int_0^\infty
\left\{\left(\bm{s}(t)\cdot \bm{I}_n\right)\int_0^t
\bm{s}(\tau)d\tau\right\}\exp\left\{-t\tau_c\right\}{\rm dt}\right\rangle_s\nonumber\\
& = & \frac{\tau_c}{4}\left\{\left(\bm{I}_{n\|}+\frac{\bm{I}_{n\perp} 2}{1+(\Omega\tau_c)^2}\right)-\frac{\left[\bm{\Omega}
\times \bm{I}_n\right]\tau}{1+(\Omega\tau_c)^2}\right\},
\label{eq:4a}
\end{eqnarray}
and finally
\begin{eqnarray}
\left\langle\frac{d\bm{I}_n}{dt}\right\rangle_s
&=&-\frac{\omega_n^2\tau_c}{4}\left\{\frac{2\bm{I}_{n\|}+\bm{I}_{n\perp}(2+(\Omega\tau_c)^2)}{1+(\Omega\tau_c)^2}+\right.\nonumber\\
&+&\left.\frac{[\bm{\Omega}\times\bm{I}_n]\tau_c}{1+(\Omega\tau_c)^2}\right\}.
\label{eq:didts}
\end{eqnarray}
The first term in the right-hand side of Eq.~(\ref{eq:didts}) contains 
the relaxation of the nuclear polarization components both parallel ($\bm{I}_{n\|}$) to and 
transverse ($\bm{I}_{n\perp}$)  to  $\bm{\Omega}$, see Eq.~(\ref{eq:didtit}). Its last term describes
the nuclear spin precession in the indirect hyperfine field, see Eq.~(\ref{eq:inrel}).

\bibliography{thesis}

\end{document}